\documentclass{PoS}
\usepackage{breqn}

\title{Update on the improved lattice calculation of direct CP-violation in K decays}

\ShortTitle{Direct CP-violation on the lattice}

\author{\speaker{Christopher Kelly}\\
	Columbia University\\
        E-mail: \email{ckelly@phys.columbia.edu}}

\author{Tianle Wang\\
	Columbia University\\
        E-mail: \email{tw2507@columbia.edu}}
        
\abstract{We discuss the status of the RBC \& UKQCD collaboration's lattice determination of $\epsilon'$, the measure of direct CP-violation in kaon decays, focusing in particular on recent developments in statistical techniques for estimating standard errors and goodness-of-fit metrics for large amounts of data that have correlations both in the temporal coordinate and also in molecular dynamics time. A key result is the formulation of a technique for determining the null distribution of a fit using a bootstrap method that is free from the usual assumptions of independence, large-$n$ and/or normal data, that can for instance be applied to compute p-values even for uncorrelated fits.}

\FullConference{37th International Symposium on Lattice Field Theory - Lattice2019\\
		16-22 June 2019\\
		Wuhan, China}

\begin{document}
\vspace{-2cm}
\section{Introduction}
\vspace{-0.2cm}
The violation of the CP symmetry in particle decays is highly suppressed in the Standard Model and therefore represents an ideal probe for new physics that may help explain the origin of the matter/antimatter asymmetry in the Universe. This direct CP-violation was initially discovered in $K\to\pi\pi$ decays, and is parameterized by a quantity $\epsilon'$. Since our publication in 2015~\cite{Bai:2015nea} of the first complete lattice calculation of $\epsilon'$ with systematically improvable errors, we have devoted much effort to improve the somewhat large statistical and systematic errors. 

In this calculation it is crucial to completely understand the $\pi\pi$ system: Not only are the energies and amplitudes required to extract the matrix elements from the lattice three-point functions, but the scattering phase shift enters into the relationship between $\epsilon'$ and the decay matrix elements, and its energy dependence is required for the Lellouch-L\"uscher finite-volume correction. 

One outstanding puzzle in our earlier work was a discrepancy of ${\sim}2\sigma$ between the measured value of the $I=0$ phase shift, $\delta_0$, and the value predicted by combining dispersion theory with experimental input~\cite{Colangelo:2001df}. A subsequent increase in statistics from 216 to 1438 configurations -- almost a factor of 7 -- served only to increase the significance to over $5\sigma$. The most likely explanation is the existence of excited-state contamination hidden beneath the rapidly growing statistical noise. This prompted us to introduce two additional $\pi\pi$ operators allowing us to perform simultaneous fits to better isolate the ground state. As reported in Ref.~\cite{twanglat2019}, applying this technique with 741 measurements appears to have resolved the discrepancy with the dispersive prediction.

A significant component of our improved analysis has been the development of advanced statistical techniques for reliably assessing standard errors and the goodness-of-fit for fits to highly correlated data with large numbers of data points and in the presence of autocorrelations within the ensemble. In this document we motivate and describe these techniques and conclude with an update on the status of the $K\to\pi\pi$ calculation to which they are also applied.
\vspace{-0.4cm}
\section{Managing autocorrelations in correlated fits}
\vspace{-0.2cm}
In order to minimize the statistical error we perform correlated fits, using the covariance matrix estimated from the data. These are performed in the context of the jackknife procedure for estimating the standard error whereby the original ensemble of size $n$ samples is {\it resampled} to form $n$ {\it reduced ensembles} of $n-1$ samples by systematically eliminating consecutive samples:
\vspace{-0.1cm}
\begin{equation}
{\bf X}_j = \left\{ x_1, x_2,\ldots, x_{j-1}, x_{j+1},\ldots x_n\right\}\,,
\vspace{-0.2cm}
\end{equation}
for $j\in\{1\ldots n\}$. The variance of the means of the reduced ensembles is related to the square of the standard error by a factor of $n-1$, and the same applies to the parameters of any subsequent fits provided the covariance matrix is also determined independently for each reduced ensemble. 

With our increased statistics we have uncovered evidence of small autocorrelation effects in our data sets, with an integrated autocorrelation time of $\tau_{\rm int}{\sim}4$ molecular dynamics (MD) time units. Na\"ively, accounting for this autocorrelation can be expected to increase our statistical error by $\sqrt{2(\tau_{\rm int}/\Delta)}\approx 1.4$, where $\Delta=4$ is the MD time separation between measurements.

\begin{figure}[tb]
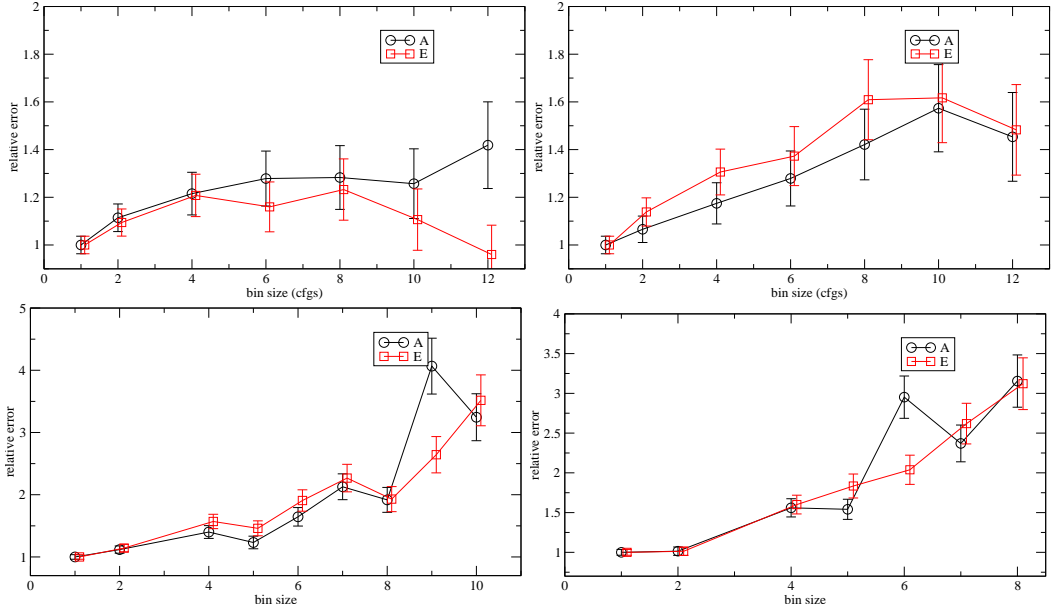

\vspace{-0.6cm}
\centering
\includegraphics[width=0.45\textwidth, trim={0.1cm 0.1cm 0cm 0.0cm},clip]{pion_bindep.eps}
\includegraphics[width=0.45\textwidth, trim={0.1cm 0.1cm 0cm 0.0cm},clip]{kaon_bindep.eps}\\
\includegraphics[width=0.45\textwidth, trim={0.1cm 0.1cm 0cm 0.0cm},clip]{plot_err_bindep_3op_3state.eps}
\includegraphics[width=0.45\textwidth, trim={0.1cm 0.1cm 0cm 0.0cm},clip]{plot_err_bindep_3op_3state_scrambled.eps}
\vspace{-0.3cm}
\caption{The error of the ground-state amplitude ($A$) and energy ($E$) relative to the unbinned error as a function of the bin size for the pion (upper-left) and kaon (upper-right) and $\pi\pi$ (lower-left). The lower-right pane shows the result of repeating the $\pi\pi$ analysis after randomly scrambling the data samples to remove autocorrelations. The error bars indicate an estimate of the error-on-the-error of $\sigma/{\sqrt{n/B}}$. \label{fig-pikbindep} }
\end{figure}

The simplest method to account for autocorrelations is to {\it bin} the data, i.e. average over $B$ consecutive measurements to obtain $n/B$ binned samples, prior to resampling. $B$ is then increased until the standard error plateaus, at which point the bins are sufficiently uncorrelated. As illustrated in Fig.~\ref{fig-pikbindep} this procedure has the expected impact upon our fits to the pion and kaon two-point functions, for which the relative error plateaus at around the predicted value of 1.4 for $B\gtrsim 6$.

\begin{figure}[tb]
\centering
\includegraphics[width=0.45\textwidth, trim={0.1cm 0.1cm 0cm 0.0cm},clip]{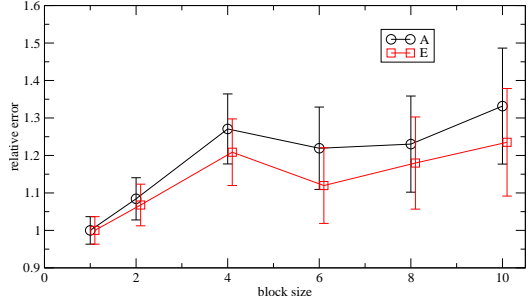}
\vspace{-0.3cm}
\caption{The error of the ground-state amplitude ($A$) and energy ($E$) relative to the unblocked error as a function of the block size for the $\pi\pi$ fit using the block jackknife technique. The error bars indicate an estimate of the error-on-the-error of $\sigma/{\sqrt{n/B}}$. \label{fig-pipibdj} } 
\vspace{-0.5cm}
\end{figure}

The above is in marked contrast to the fits to the $I=0$ $\pi\pi$ two-point function also shown in Fig.~\ref{fig-pikbindep}, whose errors increase significantly and do not appear to plateau. This behavior is explained by the nature of the correlations and the sheer size, $66\times 66$ in this case, of the covariance matrix: such a large matrix of correlated data is likely to have numerous low eigenvalues that dominate the matrix inverse. As $B$ is increased, the $n/B-1$ samples in the reduced ensemble become insufficient to resolve the correlations, resulting in ever-increasing fluctuations in these low eigenmodes between jackknife samples. This breakdown of the binning procedure is {\it unrelated} to the autocorrelations: To demonstrate this, in the same figure we repeat our analysis after randomly scrambling the data samples to destroy any autocorrelations and observe similar uncontrolled growth.

One solution is to perform a {\it block jackknife}, whereby rather than averaging over $B$ successive data prior to jackknifing, we instead generate reduced ensembles by systematically removing blocks of $B$ consecutive samples without pre-averaging:
\vspace{-0.2cm}
\begin{equation}
{\bf X}_j = \left\{ x_1, x_2,\ldots, x_{B(j-1)}, x_{Bj+1},\ldots x_n\right\}\,,
\vspace{-0.2cm}
\end{equation}
for $j\in\{1\ldots n/B\}$. As such the reduced ensembles contain $n-B$ samples with which to obtain the covariance matrix. In Fig.~\ref{fig-pipibdj} we demonstrate that this procedure results in statistical errors on the $\pi\pi$ fit parameters that now plateau as expected.

Note that the covariance matrix obtained in the block jackknife procedure does not include the effect of autocorrelations. However, as the jackknife ensures correct standard errors for any form of the covariance matrix, even a diagonal matrix for uncorrelated fits, the result is merely that we do not use the {\it optimal} matrix that formally minimizes the errors. Another effect is that the test statistic that we minimize becomes incorrectly scaled which prevents assessing the goodness-of-fit by traditional means. The procedure we will discuss in the next section circumvents this issue.
\vspace{-0.4cm}
\section{Estimating the goodness-of-fit}
\vspace{-0.3cm}

In frequentist statistics the goodness-of-fit is represented by a probability that the data is described by the model, allowing only for statistical fluctuations. Below we describe how this can be estimated correctly with only an ${\cal O}(1/n)$ bias giving a result that is free from the -- often invalid -- assumptions upon which traditional methods rely.

Any data sample $x_{\alpha i}$ at coordinate $i$ and sample index $\alpha$ can be written as
\vspace{-0.2cm}
\begin{equation}
x_{\alpha i} = f(\vec p, i) + \epsilon_{\alpha i}\,,\label{eq-samplemodelf}
\vspace{-0.2cm}
\end{equation}
where $f$ is the model function, $\vec p$ its parameters, and $\epsilon_{\alpha i}$ is a correction that may comprise both statistical fluctuations and systematic corrections if the model is imperfect. For large $n$ the population of ensemble means $\bar x_i = \frac{1}{n}\sum_\alpha x_{\alpha i}$ is normally distributed due to the central limit theorem. Therefore, in the case that the data are truly described by the model with only statistical fluctuations (the {\it null hypothesis}), this implies $\bar\epsilon_i = \frac{1}{n}\sum_\alpha \epsilon_{\alpha i}\sim {\cal N}(0,\sigma_i^2)$. 

We will use the conventional test statistic to determine the goodness-of-fit,
\vspace{-0.2cm}
\begin{equation}
q^2 = \sum_{i,j} \left(\bar x_i - f(\vec p, i)\right) ({\rm cov})^{-1}_{ij} \left(\bar x_j - f(\vec p, j)\right)\,.\label{eq-q2def}
\vspace{-0.3cm}
\end{equation}
This quantity is typically labeled $\chi^2$ but this is often a misnomer, as we will describe. For data consistent with the null hypothesis, the resulting distribution of $q^2$ over many independent experiments defines the {\it null distribution}, which also incorporates the fluctuations of the covariance matrix. 

Given a particular test statistic and the corresponding null distribution, the likelihood that a set of data is consistent with the null hypothesis can be assessed by computing the p-value:
\vspace{-0.2cm}
\begin{equation}
P(q^2) = \int_{q^2}^{\infty} dq^{\prime\,2}\ {\cal P}_{\rm null}(q^{\prime\,2})\label{eq-pvaluedef}
\vspace{-0.2cm}
\end{equation}
where ${\cal P}_{\rm null}$ is the PDF of the null distribution and $q^2$ the measured statistic. A low p-value could indicate that we were ``unlucky'', or that there is a systematic deviation from the model.

Our task is to determine the null distribution. In the limit of large $n$, the fluctuations of the covariance matrix can be ignored and the null distribution becomes the $\chi^2$ distribution for $K=T - F$ degrees of freedom, where $T$ is the number of data point and $F$ the number of free parameters. However, the results of the previous section suggest that in the case of the $\pi\pi$ two-point fits the covariance matrix fluctuates quite significantly suggesting that the $\chi^2$ distribution is not appropriate. 

If the underlying distribution is normal and the samples independent, the null distribution for finite-$n$ can be obtained analytically: it is the Hotelling $T^2$ distribution, $T^2(K, n-1)$~\cite{hotelling1931}. Of course our data are not Gaussian but are distributed according to the QCD path integral, and while numerical experiments suggest the null distribution remains similar to the $T^2$ distribution for modest excursions from normality, we cannot rely upon this fact in general. Furthermore, our experiments show that autocorrelations have a significant effect upon the shape of the distribution that can only be overcome by binning with large bin sizes, a technique inapplicable to the $\pi\pi$ fits. Nevertheless the fact that the mean of $T^2$ approaches that of $\chi^2$ like $K^2/n$, suggests that the ratio of the square of the number of degrees of freedom to the number of samples is a useful rule-of-thumb as to when one must take into account the fluctuations of the covariance matrix.

We now introduce a procedure to estimate the null distribution directly from our data via bootstrap resampling. The bootstrap generates resampled ensembles by selecting $n$ data at random (with replacement) from the original data set. This is repeated $N_{\rm boot}{\sim}1000$ times and some statistic is measured on each. As with the jackknife, the distribution of this statistic over the resampled ensembles reflects the corresponding distribution of the population, but while the jackknife distribution shares only its mean and second moment with the population distribution, the bootstrap approximates the entire distribution, in this case of $q^2$.

For a bootstrap resampled ensemble $b\in \{1\ldots N_{\rm boot}\}$ we define,
\vspace{-0.25cm}
\begin{equation}
\bar x^b_i = \bar x_i + \bar\epsilon_i^{\prime\,b}\label{eq-bootstrpdev}
\vspace{-0.25cm}
\end{equation}
where $\bar x^b_i$ is the mean of the resampled ensemble and $\bar x_i$ that of the original ensemble. The bootstrap analogue of the central limit theorem implies that over the bootstrap ensembles, 
\vspace{-0.2cm}
\begin{equation}
\bar\epsilon_i^{\prime\,b}{\sim}{\cal N}(0,\sigma_i^2)\label{eq-bootstrpcnlm}
\vspace{-0.2cm}
\end{equation}
up to a bias typically ${\cal O}(1/n)$. To a good approximation the distribution of covariance matrices over bootstrap ensembles also reflects that of the population.


While the bootstrap means are distributed about the original ensemble mean, they are not typically distributed about the {\it model} and therefore do not satisfy the null hypothesis (even if the original ensemble is consistent), due to the presence of the shift $\bar\epsilon_i$ in Eq.~\ref{eq-samplemodelf}. However we can {\it impose} the null hypothesis on the bootstrap ensembles by applying a {\it recentering}:
\vspace{-0.2cm}
\begin{equation}
\bar x^b_i \to \bar x^{\prime\,b}_i = \bar x^b_i + f(\vec p, i) - \bar x_i = f(\vec p, i) + \bar\epsilon_i^{\prime\,b}\,,
\vspace{-0.2cm}
\end{equation}
where $\vec p$ are the parameters obtained from the fit to the original ensemble and in the second equation we have applied Eq.~\ref{eq-bootstrpdev}. Due to Eq.~\ref{eq-bootstrpcnlm} the recentered data now satisfy the null hypothesis defined above. As the shift in the means is independent of the sample index $\alpha$, the bootstrap distribution of covariance matrices -- which reflects that of the population -- is unaffected. We can therefore obtain the corresponding null distribution by minimizing $q^2$ upon each recentered bootstrap ensemble.

This second stage of minimization is essential to obtain the null distribution with the correct number of degrees of freedom: The $F$ conditions for the minimum define an $F$-dimensional subspace in which the contributions to $q^2$ are exactly zero. As such $q^2$ arises entirely from the projection of the data and model into the $K$-dimensional subspace orthogonal with respect to the inner product defined by the covariance matrix, and this subspace depends on the covariance matrix and the fit parameters which both vary between bootstrap ensembles.

We can compute the p-value for the fit to our original data set by approximately integrating the empirical null distribution per Eq.~\ref{eq-pvaluedef}. A simpler approach is to sort the bootstrap values of $q^2$ in ascending order and find the closest value of $q^2$ within the resulting array. If this closest value corresponds to array element $i$, then the p-value is, to a good approximation, $p_i = \frac{N_{\rm boot}-i-1}{N_{\rm boot}}$.

Of course, as with most resampling techniques there is a finite-$n$ bias. We intend to investigate whether repeating the analysis using one or more bootstrap ensembles in place of the original ensemble allows the size of the bias to be estimated.

\begin{figure}[tb]
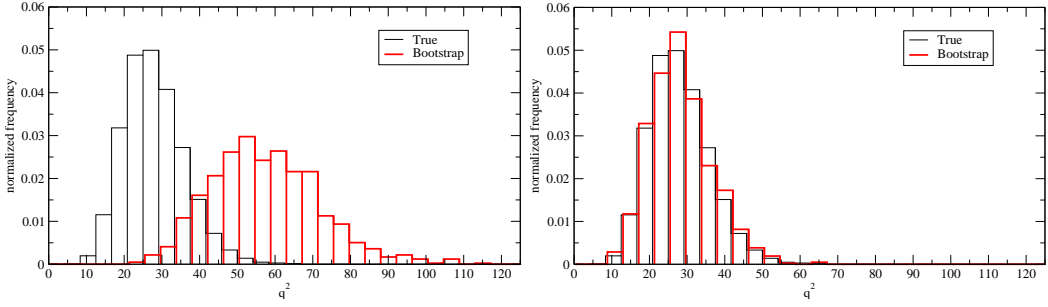

\vspace{-0.5cm}
\centering
\includegraphics[width=0.45\textwidth, trim={0.1cm 0.1cm 0cm 0.0cm},clip]{hist_q2_uncorrected.eps}
\includegraphics[width=0.45\textwidth, trim={0.1cm 0.1cm 0cm 0.0cm},clip]{hist_q2_corrected.eps}
\vspace{-0.3cm}
\caption{For $n=400$ samples of data generated according to the form $N( e^{-0.1t} + 0.5 e^{-0.5t})$ with $N{\sim}{\cal N}(1, 0.1^2)$ and $L_T=30$ we compute the true null distribution by fitting to the form $A_1 e^{-E_1} + A_2 e^{-E_2}$ for many independent experiments. In the left pane the resulting histogram of $q^2$ is compared to the bootstrap estimate without recentering, and in the right pane with the recentering applied. \label{fit-gaussrecenter} }
\end{figure}

\begin{figure}[tb]
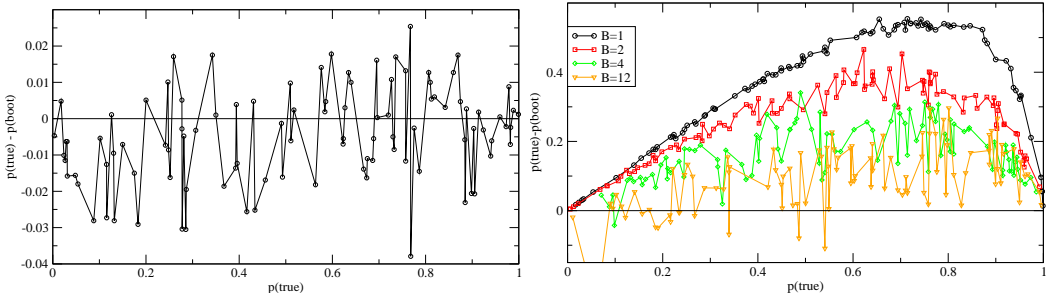

\vspace{-0.2cm}
\centering
\includegraphics[width=0.45\textwidth, trim={0.1cm 0.1cm 0cm 0cm},clip]{lognorm_diff_ind.eps}
\includegraphics[width=0.45\textwidth, trim={0.1cm 0.1cm 0cm 0cm},clip]{lognorm_0_0pt7_blockdep.eps}
\vspace{-0.3cm}
\caption{For $n=400$ samples of data generated according to the form $N( e^{-0.1t} + 0.5 e^{-0.5t})$ with $N{\sim}{\rm Lognormal}(0,0.7^2)$ and $L_T=30$ we compare the p-value predicted by the bootstrap to the true p-value for correlated fits to 100 independent experiments. Differences $>0$ indicate an underestimation of the p-value and the fluctuations give some indication of the bias. In the left pane we show the comparison for independent samples. In the right pane we repeat the analysis for autocorrelated samples with $\tau_{\rm int}\approx 32$ and configuration separation 32, using NBB (see below) and varying the block size. \label{fit-lognormcorr} }
\vspace{-0.1cm}
\end{figure}

\begin{figure}[tb]
\centering
\includegraphics[width=0.45\textwidth, trim={0.1cm 0.1cm 0cm 0cm},clip]{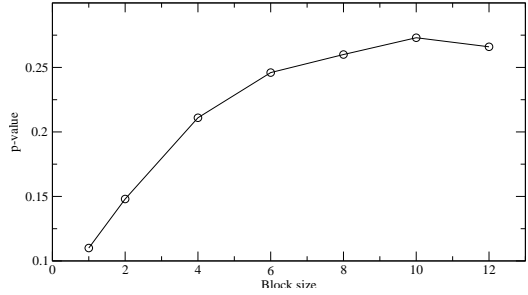}
\vspace{-0.3cm}
\caption{The dependence of the bootstrap p-value for the $\pi\pi$ fits as the block size is varied.\label{fig-pipibdep} }
\vspace{-0.3cm}
\end{figure}

In Fig.~\ref{fit-gaussrecenter} we demonstrate the recentering technique as applied to the $q^2$ distribution of Gaussian data, and in the left pane of Fig.~\ref{fit-lognormcorr} show that the bootstrap technique also applies to heavily log-normal, uncorrelated data for which the best analytic approach -- the Hotelling $T^2$ distribution -- is inapplicable. The latter also shows evidence of the expected bias, which results in a disagreement between the predicted and true p-value on the order of a few percent for 400 configurations. 

This technique can also tolerate autocorrelations by replacing the na\"ive bootstrap with a {\it block bootstrap} variant. For simplicity we use the {\it non-overlapping block bootstrap} (NBB)~\cite{nbbcarlstein}, which is the bootstrap analogue of the block jackknife: rather than selecting individual data from the original ensemble, one selects contiguous, non-overlapping blocks of size $B$ at random that are laid down in order to produce an ensemble of size $n$ (assuming $B$ is a divisor of $n$). In the right pane of Fig.~\ref{fit-lognormcorr} we repeat the demonstration for log-normal data generated using the Metropolis method to introduce autocorrelations. We observe the predicted and true p-values converge as $B$ is increased, although this also appears to increase the fluctuations resulting from the finite-$n$ bias. 

The bootstrap technique is very general and does not require assumptions as to the form of the covariance matrix. As such it will remain applicable for the incorrectly-scaled $q^2$ resulting from not incorporating the effects of autocorrelations in the covariance matrix under the block jackknife procedure of the previous section. It also can be used to obtain the goodness-of-fit for {\it uncorrelated fits} which is not possible using traditional techniques. 

In Fig.~\ref{fig-pipibdep} we apply the procedure to our $\pi\pi$ two-point fits and observe the p-value plateaus at ${\sim}0.3$, which is significantly higher than the values of 0.04 and 0.1 predicted if we neglect autocorrelations and assume the $\chi^2$ and $T^2$ distributions, respectively, and implies a good fit.
\vspace{-0.3cm}
\section{Conclusions}
\vspace{-0.2cm}

In this document we introduce techniques that allow for the more reliable determination of the standard error and the goodness-of-fit when fitting to large amounts of correlated data in the presence of autocorrelations. For the former we demonstrate the breakdown of the traditional binning approach in the form of uncontrolled error growth as the bin size is increased, and introduce the block jackknife technique to circumvent this issue. For the goodness-of-fit we discuss how the conventional techniques of computing the p-value via the $\chi^2$ or Hotelling $T^2$ distribution are inapplicable for non-normal data with autocorrelations, and introduce a bootstrap approach that is free from assumptions and is correct up to a ${\cal O}(1/n)$ bias. 

Applied to the $\pi\pi$ fits we demonstrate both improved control over the statistical error estimate and a larger, more reliable p-value. These new, multi-operator fits also appear to resolve the discrepancy in the $I=0$ $\pi\pi$ scattering phase shift between our 2015 result and the dispersive prediction through much better control over the excited state contamination. We are presently in the process of finalizing this analysis and the related analysis of the $K\to\pi\pi$ matrix elements and $\epsilon'$.

\end{document}